\documentclass[aps,prd,twocolumn,groupedaddress]{revtex4-1}

\usepackage{dsfont}
\usepackage{natbib}
\usepackage{amssymb}
\usepackage{graphicx}
\usepackage{amsmath}
\usepackage{url}
\usepackage{epstopdf}
\usepackage{color}

\def\p {\partial}

\def\be {\begin{equation}}
\def\ee  {\end{equation}}
\def\bea {\begin{eqnarray}}
\def\eea {\end{eqnarray}}
\def\nn {\nonumber}

\begin{document}

\title{Ground state of the Universe and the cosmological constant.\\A nonperturbative analysis}

\author{Viqar Husain}
\email{vhusain@unb.ca}
\affiliation{Department of Mathematics and Statistics, University of New Brunswick, Fredericton, NB, Canada E3B 5A3} 

\author{Babar Qureshi}
\email{ babar.qureshi@lums.edu.pk}
\affiliation{Department of Physics, LUMS School of Science and Engineering, Lahore, Pakistan} 

\date{\today}

\begin{abstract}

The physical Hamiltonian of a  gravity-matter system depends on the choice of time, with the vacuum naturally identified as its ground state.  We study the expanding universe  with scalar field in  the volume time gauge.  We show that   the   vacuum energy density computed  from  the resulting  Hamiltonian  is a non-linear function of the cosmological constant and time.   This result provides  a new perspective on the relation between time,  the cosmological constant,    and  vacuum energy.

\end{abstract}

%\keywords{}

\maketitle

One of the outstanding problems in fundamental physics is that of the cosmological constant (CC). The problem arises in the context of  quantum field theory  (QFT) on a fixed background spacetime, which is usually taken to be flat \cite{Weinberg:1988cp,Carroll:2000fy,Rugh:2000ji,Burgess:2013ara}, or otherwise has a high degree of symmetry. The symmetry  includes  a global notion of time specified as a timelike Killing vector field. The dynamics of the gravitational field is included only in so far as it is viewed as a spin two field on the specified background; back reaction of  quantum fields on spacetime is typically excluded.  

QFT on a fixed background spacetime may be viewed as the leading order  term coming from  the semi-classical approximation defined by the equation  
\be
 G_{ab} + \Lambda g_{ab} = 8\pi  G \langle \psi|  \hat{T}_{ab}(\hat{\phi}, g) | \psi  \rangle, \label{semicl}
\ee
where $\Lambda$ is the (bare) cosmological constant. As written, this hybrid classical-quantum equation is ambiguous. To make it more precise we require (i) a quantization of the matter field $\phi$ on   a general background $g_{ab}$, (ii) a suitably regularized self-adjoint operator $\hat{T}_{ab}$, and lastly (iii) computation of the expectation value of  $\hat{T}_{ab}$ in some choice of matter vacuum state $|\psi\rangle$.  This would  give the tensor
\be
 \tilde{T}_{ab}^\psi(g)  \equiv \langle \psi|  \hat{T}_{ab}(\hat{\phi}, g) | \psi  \rangle,
\ee
as the effective stress-energy tensor associated to the state $|\psi\rangle$, and hence a precise meaning for the r.h.s. of  eqn. (\ref{semicl}). One can then proceed to solve this equation for the ``semi-classical" metric $g_{ab}$.  

Although there is a large literature \cite{Parker:2009uva} on computations of  the r.h.s for a given spacetime, the calculation of a semiclassical metric has not been carried to satisfactory  completion, even for  spacetimes  with isometries. In fact the equation itself has been questioned \cite{Page:1981aj}.   Nevertheless, an attempt to produce a self-consistent solution by expanding the metric and state as  
\bea
g_{ab} &=& \eta_{ab} + \epsilon h^{(1)}_{ab} + \epsilon^2 h^{(2)} + \cdots  \nn\\
|\psi\rangle &=& |0\rangle + \epsilon |\psi^{(1)} \rangle  + \epsilon^2 |\psi^{(2)} \rangle + \cdots
\eea
($\epsilon = m/m_P$) leads to $0th.$ order to 
\be
\Lambda\eta_{ab} = 8\pi G \langle 0| T_{ab} | 0\rangle. 
\ee 
This equation  forms the basis of the connection between vacuum energy density $\rho_{vac}$ and $\Lambda$, specifically  the broadly accepted linear relationship  
\be
\rho_{vac}  = \frac{\Lambda}{8\pi G}.
\ee
  It   leads to the cosmological constant problem via the elementary  evaluation
  \be
\rho_{vac}=\frac{E}{V} =   \int_0^{k_p}  \frac{d^3k}{(2\pi)^3} (\hbar k) = \frac{\hbar}{8\pi^2} k_p^4,
\ee
where $k_p$ is a Planck scale cutoff.   This huge quantity is often compared to the observed WMAP value  
\be
\Lambda= 1.27 \pm 0.07 \times 10^{-56} \text{ cm}^{-2} \label{wmapL}
\ee
$( \sim  3.2 \times 10^{-122}\   l_P^{-2})$ as a significant failure of theory.

A more sophisticated  argument presents this issue as a problem coming  from running scales in the theory. Assuming a fixed background that defines energy $k$,  the regulated vacuum energy density computed from $\langle 0| \hat{T}_{ab}|0\rangle$ is expected  (on dimensional grounds) to be of the form
\be
\rho_{vac} = M^4 f(k; g_1,g_2,\cdots) = \frac{\Lambda(k)}{8\pi G(k)}.  
\ee
where $f$ is a function of energy scale $k$,   matter coupling constants $g_1,g_2\cdots$, and  some natural mass scale $M(k)$. The first equality comes from field theory, and the second from semiclassical general relativity.  (This  expression assumes the usual  linear dependence of energy density on $\Lambda$, and is  observer 4-velocity $v^a$ dependent:  $\displaystyle \rho =  v^a v^b\langle 0| \hat{T}_{ab}|0\rangle$, unless there is a preferred timelike vector field specified by a spacetime isometry.)  In this setting there are two ways to state the  CC  problem: (i)  it arises from the first equality due to the factor $M^4$ which gives a very large energy density even well below the Planck energy, for example  for proton mass or $\Lambda_{QCD}$, or (ii)  it arises from the second equality as a fine tuning problem; at low energies (1 meter to a few astronomical units) where 
$G$ and $\Lambda$ are observed to be constant, the corresponding dimensionless parameters flow canonically as $\lambda(k) = \Lambda /k^2$ and $g(k) = G k^2$. Thus the low energy renormalization group trajectory must be a hyperbola $\lambda(k) g(k)=$ constant, which reflects a fine tuning of the initial conditions for the flow  \cite{Reuter:2004nx}.  

The field theory problem may be due to the fact that the function $f$ is usually computed in perturbation theory. A counterpoint is provided by a recent non-perturbative calculation in the Gross-Neveu model, which  suggests that, non-perturbatively, $f$ is a non-analytic function of the coupling constant that suppresses $\rho_{vac}$ at low energy \cite{Holland:2013xya}.  

 We question the basis of formulating the CC problem to first order in the semiclassical setting, and argue that in a non-perturbative quantum approach in which gravitational degrees of freedom are treated as a part of the dynamics, either the problem does not arise, or that its manifestation is substantially different from that coming from the usual arguments.    
 
 We take the view that to meaningfully  talk  about a vacuum, we need a physical Hamiltonian for the full gravity-matter system. This in turn requires a global notion of time in the context of a generally covariant theory. Hence {\it there is a connection between non-perturbative vacuum energy, the cosmological constant,   and  a global  time variable. } However as there is no ``solution to the problem of time" in quantum gravity,    one might  impose a plausible  time gauge, or use some other suitably defined ``relational time."  We will use geometry degrees of freedom  to fix time gauge and derive the corresponding physical Hamiltonian. The spectrum of the corresponding  operator then gives a formula for the vacuum energy density.
 
  The suggestion that quantum gravity might play a role in its resolution is not new; see eg. \cite{Witten:2000zk} in the context of string theory, \cite{Husain:2009cf} in the Hamiltonian context which is developed further here, and a semiclassical  approach using Regge calculus \cite{Mikovic:2014opa}.

With this summary and  context, we begin with the 3+1 Arnowitt-Deser-Misner (ADM) Hamiltonian for Einstein gravity and minimally coupled to a massive scalar field 
\be
S = \int d^3x dt \left( \pi^{ab}\dot{q}_{ab} + P_\phi\dot{\phi} - NH -N^a C_a  \right),
\ee
where $(q_{ab},\pi^{ab})$ and $(\phi,P_\phi)$ are the ADM gravitational and scalar field phase space variables, $N, N^a$ are the lapses and shift variables, and 
\bea
H &=& \frac{1}{\sqrt{q}} \left(  \pi^{ab}\pi_{ab} - \frac{1}{2} \pi^2 \right) +  \sqrt{q}( \Lambda - R) + H_\phi  \label{Hcons}\\
 C_a &=& D_b\pi^b_{\ a} + P_\phi \p_a\phi, \\
 H_\phi &=& \frac{1}{2} \left( \frac{ P_\phi^2}{\sqrt{q}}  + \sqrt{q} q^{ab} \p_a\phi\p_b\phi  + \sqrt{q} m^2 \phi^2\right). 
\eea
are respectively the Hamiltonian and diffeomorphism constraints, and the scalar field Hamiltonian density. (We work in geometric units where $G=\hbar=c=1$, and reintroduce these constants in the final result.)

From this starting point, our goal is to calculate the vacuum energy density of the scalar field $\rho_{vac}(\Lambda,m)$ derived from the physical Hamiltonian associated to the volume time gauge in a cosmological setting. We do this first  in the homogeneous (zero mode) setting to illustrate the argument, and subsequently  generalize it to include  all matter modes.  
  
The flat homogeneous model is derived by the parametrization
\be
q_{ab} = a^2 e_{ab},\ \ \ \ \  \pi^{ab} = \frac{p_a}{6a}\  e^{ab}
\ee  
where $e_{ab}=\text{diag}(1,1,1)$. Substituting this into the constraints and ADM action gives the reduced theory
\be
S = V_0 \int dt \left(\dot{a}p_a + \dot{\phi}  P_\phi - N H   \right), \label{red-act}
\ee
where 
\be
 H = -  \frac{p_a^2}{24a} + a^3\Lambda + \frac{1}{2} \left( \frac{ P_\phi^2}{a^3}  + a^3 m^2 \phi^2\right).\label{frwH}
 \ee
 The last equation is  obtained from substituting the reduction ansatz into the Hamiltonain constraint (\ref{Hcons}), and $V_0$ is an unphysical  coordinate volume.  
The reduced  action is  invariant under the scale transformations
\be
(V_0, a, p_a, \phi,P_\phi) \rightarrow \left(\lambda^3 V_0,  \frac{a}{\lambda}, \frac{p_a}{\lambda^{2}}, \phi, \frac{P_\phi}{\lambda^{3}}   \right) \label{scale}
\ee 
 
 At this stage we fix ``physical volume time" gauge \cite{Hassan:2014sja}  by setting 
\be
t =   \int d^3x \sqrt{q} =  V_0a^3. \label{tgauge}
\ee
 We note that this is both scale invariant (\ref{scale}) and  second class with the Hamiltonian constraint, as required of an adequate gauge fixing. It  is also a physically natural time in the context of an expanding cosmology. Although we do not require it here, the lapse function corresponding to this canonical time gauge is given by the requirement that the gauge be preserved under evolution.
 This gives  
 \be
 1= \{V_0 a^3, NH  \} = -\frac{V_0Nap_a}{4}  \implies N = -\frac{4}{V_0 ap_a}.   
 \ee
We note that this lapse is  invariant under the transformation (\ref{scale}), as it should be. 

This gauge  condition, together with the solution of the Hamiltonian constraint,  eliminate the variables $(a,p_a)$, leaving a theory for the scalar field variables evolving with respect to this time. 
The gauge fixed canonical action is obtained by substituting (\ref{tgauge}) and the solution of the Hamiltonian constraint 
\be
p_a^2 = 24 \left[   a^4\Lambda + \frac{a}{2} \left( \frac{ P_\phi^2}{a^3}  + a^3 m^2 \phi^2\right)      \right]_{|V_0a^3 =t}
\ee
into the action (\ref{red-act}). We  choose the root that gives positive energy density. 

It is useful to write the gauge fixed action  using the scale invariant variables $p_\phi := V_0P_\phi$  and $t$. This  gives  
\be
S^{GF} = \int dt \  \left(   \dot{\phi}  p_\phi  - H_P \right),
\ee
where  
\be
H_P=   \sqrt{   \frac{8}{3}\left( \Lambda +  \frac{p_\phi^2}{2t^2} + \frac{1}{2}m^2 \phi^2    \right)        }  \label{Hp}
\ee
The energy density derived from this Hamiltonian is 
\be
\rho = \frac{H_P}{V_0a^3} = \frac{H_P}{t}, \label{rho}
\ee
since $V_0a^3$ is the physical volume (which is also  the chosen time gauge).  We note that this physical quantity does not depend on $V_0$. 

To find the eigenvalues  of this density operator we recall that for any operator $\hat{A}$ with a positive spectrum $a_n$, the spectrum of the square root operator $\sqrt{\hat{A}}$,  is  $\sqrt{a_n}$.  In our case the argument of the square root in (\ref{Hp}) is  a shifted  harmonic oscillator with time dependent mass and frequency.  Therefore we can solve  the eigenvalue problem for the density operator $\hat{\rho}$  
\be
\hat{\rho} \ \psi = \rho_n \psi 
\ee
by treating $t$ as a parameter. This gives the exact spectrum
\be
\rho_n =  \frac{m_p^2}{t} \sqrt{   \frac{8}{3} \left[ \Lambda  + \left(n+ \frac{1}{2} \right)  \frac{m}{t} \  \right] }, \label{rhon}
\ee
where $n=0, 1, \cdots $, and we have reintroduced the Planck mass, with $\Lambda, m,$ and $t$ specified in Planck units.  

Let us note that this energy density operator may also be used to set up the time dependent Schrodinger equation, specify an initial state, such as the $n=0$ state, and evolve it to the present time. In general, such an  evolved state may be approximated  by a finite linear combination of the instantaneous energy eigenbasis $|\psi_n\rangle $ of the density operator,
\be
|\Psi(t)\rangle = \sum_{n=0}^N  c_n(t) |\psi_n(t)\rangle.
\ee
Now for our purpose, which is to obtain a relationship between energy density and cosmological constant, we would need to evaluate the expectation value of the density operator  in this state 
\be
\langle \Psi(t) | \hat{\rho} (t) |\Psi(t)\rangle = \sum_{n=0}^N  c_n(t) \rho_n(t). \label{state}
\ee
However this is not necessary to make the central point of the paper, as we now show.

The expression gives a non-perturbative quantum energy density of the scalar field with respect to the volume time gauge (\ref{tgauge}).   The eigenvalue  $\rho_n(t)$ in this formula has some interesting features:  (i)  it depends  only on variables invariant under the scale transformations  (\ref{scale}), (ii) there is a  square root  arising from the fact that all terms in the Hamiltonian constraint are quadratic in momenta, (iii) the energy density is  not linear in $\Lambda$, (iv) there is  a time factor suppression which for large times gives 
\be
\rho_{vac} \equiv m_P^2\  \sqrt{\frac{8\Lambda}{3t^2}},  
\ee 
independent of $n$. 

These features are not what are expected from the usual flat space arguments for matter vacuum energy density, where this  density is linear in $\Lambda$ and time independent.  
The last formula may be viewed as a prediction for the (zero mode) quantum vacuum energy density  of the scalar field in an FRW universe, since this factor comes out of the sum (\ref{state}) for late times. (We note that at each $t$ the state lives in the instantaneous Hilbert space at that time, so the remaining sum adds to unity. )

A numerical estimate of $\rho_{vac}$ using known cosmological parameters may be computed using the measured WMAP value for $\Lambda$ in eqn.  (\ref{wmapL}) and the present age of the universe $t = 10^{61}t_P$, ($t_P$ = Planck time). This gives  
 \be
\rho_{vac} \sim 5\times 10^{-129} \rho_{P} = 2.5\times 10^{-32}\  \text{Kg}/\text{m}^3, 
\ee
where $\rho_P =m_P/l_P^3$ is the Planck density. (We note that experiments such as WMAP measure cosmological model parameters such as $\Lambda$; implications for vacuum energy density are then derived from  theoretical models. That is, there is no direct measurement of the energy density in a box of empty space.) 

In summary to this point, we have seen that the time dependence in (\ref{rhon}) has its origin in factors of $\int d^3x \sqrt{q}=V_0a^3 =t$; the overall factor $1/t$ comes from converting the Hamiltonian scalar density (of weight one) to a scalar, and the factor in the oscillator frequency comes from the $\sqrt{q}$ terms  in the matter Hamiltonian. 

The semiclassical calculation of energy density is via $\rho = \langle 0|T_{ab} -\Lambda g_{ab} |0\rangle v^av^b \equiv \rho_\phi + \rho_\Lambda$ for an observer with four velocity $v^a$. How is this to be compared with our result eqn.  (\ref{rhon})?  It is clear that the latter is additive in the contributions from matter and $\Lambda$, whereas our result (\ref{rhon}) is not. It shows that  imposing a time gauge, solving the Hamiltonian constraint,  and then diagonalizing the resulting physical Hamiltonian is an entirely  different process from QFT on a fixed background, and yields substantially different results. 

The setting we have discussed so far is obviously limited  without a field theory extension to include all matter modes. This requires inclusion of inhomogeneities in the matter and metric degrees of freedom. We now turn to this.  We will see that  the main features of the energy density formula  (\ref{rhon})  -- explicit time dependence and the square root -- remain unaltered.  

 We follow a hamiltonian approach similar to that  developed in \cite{Langlois:1994ec}, where the scalar field and metric perturbations are expanded in Fourier modes, and the Hamiltonian constraint is treated to second order in the perturbations. The resulting theory describes the dynamics of the  gravity phase space variables $(a,p_a)$, and the scalar field and metric perturbation Fourier mode pairs $(\phi_k, p_k)$ and  $(\delta q_{ab}^k, \delta \pi^{ab}_k)$;  the mode decomposition is defined using the global chart on homogeneous space slices: 
 \bea
  \phi({\bf x},t) &=&   \sum_{\bf k}    \phi_{\bf k} (t)  e^{  i {\bf k}  \cdot {\bf  x}   }   \nn\\
 P_\phi( {\bf x}, t) &=& \sum_{\bf k}   P_{\bf k} (t)  e^{i{\bf k}\cdot {\bf x}}.
 \eea
  This gives  
 \be
 H_\phi =  V_0\sum_{\bf k}  \left(   \frac{P_{\bf k}^2}{2a^3} +  \frac{a|{\bf k}|^2}{2}\phi_{\bf k}^2 + \frac{a^3 m^2}{2} \phi_{\bf k}^2    \right)  \label{Hk}
 \ee
 after a suitable mode relabelling. The Fourier modes so defined satisfy the equal time  Poisson bracket  
 \be
 \{\phi_{\bf k}(t), P_{\bf k'}(t) \}= \delta_{{\bf k},{\bf k}' } .
 \ee
 With this decomposition we define a Hamiltonian system by the phase space variables $(a,p_a)$ and $(\phi_{\bf k}, P_{\bf k})$ and action  
 \be
 S = V_0 \int dt\left(\dot{a}p_a + \sum_{\bf k} \dot{\phi}_{\bf k} P_{\bf k}  - N H \right). \label{kact}
 \ee
 where  
 \be
 H \equiv -  \frac{p_a^2}{24a} + a^3\Lambda + \bar{H}_\phi =0, 
 \ee
 with $\bar{H}_\phi = H_\phi/V_0$ from  (\ref{Hk}).   This Hamiltonian constraint generalizes (\ref{frwH}) to include an infinite number of degrees of freedom. 

 This system is not exactly that obtained from metric and matter perturbations in \cite{Langlois:1994ec}; in particular it does not include the spatial diffeomorphism constraint, which would impose further conditions between the matter and gravity modes. Nevertheless it is a consistent model that has the main features of interest for our purpose, which is to investigate the  vacuum energy density  of a matter-gravity system with an infinite number of degrees of freedom.  We note also that  the action (\ref{kact}) has the scaling invariance (\ref{scale}) with $(\phi, P_\phi)$ replaced by their Fourier modes $(\phi_{\bf k}, P_{\bf k})$.
 
Proceeding as for the homogeneous case, let us  fix the (scale invariant) time gauge  (\ref{tgauge}) and solve the Hamiltonian constraint to eliminate $(a, p_a)$. Using the scale  invariant momentum $p_{\bf k } \equiv V_0 P_{\bf k}$, the gauge fixed action for the matter modes is 
\be
S^{GF} = \int dt \sum_{k} \left( \dot{\phi}_{\bf k} p_{\bf k} - H_P \right),
\ee    
\be
H_P =  \sqrt{   \frac{8}{3}\left( \Lambda +   \sum_{\bf k} \left[  \frac{p_{\bf k}^2}{2t^2} + \frac{1}{2} \left(t^{-\frac{2}{3}} |\bar{\bf k}|^2  + m^2 \right) \phi_{\bf k}^2             \right] \right)   }. 
\ee
where $\bar{\bf k} = {\bf k} V_0^{\frac{1}{3}}$ is the (scale invariant)  wave vector. Upon quantization the corresponding operator  has the spectrum

 \be
E_n =\sqrt { \frac{8}{3} \left[ \Lambda + \sum_{\bf k}   \left( n+ \frac{1}{2} \right) \omega_{\bf k} (t)  \right]  },\label{En}
\ee
\be
  \omega_{\bf k}(t)  = \frac{1}{t}  \sqrt{  t^{-\frac{2}{3}} \bar{{\bf k}}^2 + m^2}. \label{freq}
 \ee 
 
To find the  vacuum energy density of the matter modes  we again set $n=0$, and consider the massless case $m=0$ for simplicity.  The $\sum_{\bf k}$ is a sum over  comoving modes, which is evaluated by converting the sum to an integral in the usual way with a $k-$space volume $d^3k$: 
\be
\sum_{\bf k} \omega_{\bf k} \rightarrow  \frac{ 1}{t^{\frac{4}{3}} } \int_0^{\bar{k}_p} \frac{d^3{\bar k}} {(2\pi)^3} \ \bar{k}.
\ee
Restoring factors of Planck mass, this gives for vacuum energy density   the result 
\be
\rho_{vac} = \frac{E_0}{t} = \frac{\rho_P}{\bar{t}}  \sqrt{   \frac{8}{3} \left(\Lambda l_p^2 +    \frac{1}{8\pi^2 \bar{t}^{\frac{4}{3} } } \right)} , \label{rho0-matter}
\ee
where $\bar{t}=t/t_P$. The overall factor is the same as that for the homogeneous case. The peculiar time factor multiplying the second term in the square root comes from  the   mode frequency (\ref{freq}), which in turn has its origin in the scalar field gradient term $\displaystyle \sqrt{q}q^{ab}\p_a\phi\p_b\phi \rightarrow a^3 |{\bf k}|^2/a^2$.   

It is apparent that the general features of the homogeneous case, the square root and explicit time dependence  are still present. We may again compute a numerical  estimate
for the vacuum energy by substituting on the r.h.s.  of (\ref{rho0-matter}) the present age of the universe  $\bar{t} = 10^{61}$ and the $\Lambda$ value from (\ref{wmapL}). This gives 
\be
\rho_{vac} \sim  \rho_P \times  10^{-103} =5\times  10^{-7} {\text Kg}/{\text m}^3.
\ee
This formula makes clear that the present vacuum energy density with the  global choice of volume time  is far smaller than the huge value from standard arguments. It shows that there is no cosmological constant problem. 

Let us summarize our main result. We find  for the non-perturbative matter-gravity system in the cosmological context that  the physical Hamiltonian  (i) is not a linear function of $\Lambda$, (ii)  is explicitly time dependent, and (iii) yields the explicit formula (\ref{rho0-matter}) for vacuum energy density.   A numerical evaluation of this density shows that the vacuum energy problem is absent  due to  the time suppression factor.   Beyond these details, our general argument reveals  that there is an intimate connection between time, vacuum energy and the cosmological constant, which is revealed by extracting the physical Hamiltonian for a matter-gravity system is a physically reasonable time gauge.  (For negative cosmological constant our results do not apply above a critical time value. This means that the volume time gauge does not provide a useful foliation.)

In closing we provide several comments on our approach,  pointing out what  we think are generic features  and  what are limitations which merit further work. 

\noindent (i) In  FRW cosmology the matter energy density is identified as the right hand side of the Friedmann equation. This is fine for classical theory, but a non-perturbative quantum theory requires a physical Hamiltonian for the full matter-gravity system before one can talk about the true vacuum.

\noindent (ii) Our  approach does not address the question of why the observed cosmological constant is so small. But it does address the problem of the relation between  vacuum energy density and the cosmological constant; this we show is time dependent and non linear.

\noindent(iii) The functional form of the physical Hamiltonian, and hence the vacuum energy density is dependent on the time gauge. The square root and time dependent physical Hamiltonian are a common feature of canonical time gauge fixing. This is because the  Hamiltonian constraint is quadratic in momenta for usual matter fields (see \cite{Husain:2011tk} for an unusual exception). As a result one ends up solving at least a  quadratic equation for the  momentum conjugate to the chosen time variable. 

\noindent(iv) Our results are derived in only one time gauge in the setting of FRW cosmology with perturbations. Although this is observationally relevant, for more general metrics it is not possible to use  volume time because  it does not provide a complete time gauge fixing. The general problem is more challenging. It requires fixing a suitable  local matter or geometry scalar  as time, and deriving the corresponding Hamiltonian density.  The latter may not be a simple function, and the spectrum problem correspondingly difficult.

\noindent(v) Beyond the homogeneous case,   our development uses the fixed volume time gauge from the background to define the physical Hamiltonian of matter perturbations.  The spectrum of this Hamiltonian provides only the energy part of the semiclassical equation (\ref{semicl}).  The  pressures can be computed, and would come from   analyzing the  spatial diffeomorphism constraint 
$D_b\pi^{ab} = j^a(\phi)$  to leading order beyond the homogeneous approximation (where this constraint  is trivially satisfied).  This would be among  the necessary steps for developing a canonical semiclassical approximation using our approach as a starting point.  

\noindent(vi)  We used a Planck scale cutoff in deriving the vacuum energy density (\ref{rho0-matter}). Our justification of this is the same as that in the usual treatment because the scalar perturbations are effectively being treated on the FRW background. That is,  it is not yet full quantum gravity. But the novel feature in the formula, unlike the flat space case, is the time factor suppression of this term in   (\ref{rho0-matter}), which leaves the $\Lambda$ factor as the dominant one at late time. 

\noindent  (vii) What becomes of the ``low energy" CC problem in a small patch of spacetime where there is a local timelike Killing vector field? This  local Minkowski  time is obviously  fine  for short timescale particle physics during which  the universe does not expand much.  But our approach and results suggest  that it is not useful to pose  questions such as  ``does the vacuum gravitate"   in a local flat patch of a cosmological spacetime. 

 \bigskip
 
\noindent \underbar{Acknowledgements:} The work  was supported in part by a  grant from the Natural Sciences and Engineering Research Council of Canada. We thank Jack Gegenberg, Tim Koslowski,  Sanjeev Seahra and Jon Ziprick for discussion and comments on the manuscript.

\bibliography{E-Lambda}

\end{document}